\documentclass[prd,aps,nofootinbib,onecolumn,showpacs,12pt]{revtex4-1}
\usepackage{graphicx,epsfig}

\usepackage{epsf}
\usepackage{graphicx}

\begin{document}

\title{  \bf   Masses and decay constants of bound states containing   fourth family quarks from QCD sum rules}
\author{ V. Bashiry$^{\dag1}$, K. Azizi$^{\ddag2}$,  S. Sultansoy$^{*3}$ \\
$^{\dag}$Engineering Faculty, Cyprus International University,
Via Mersin 10, Turkey\\
$^{\ddag}$Physics Division,  Faculty of Arts and Sciences, Do\u gu\c s
University,
 Ac{\i}badem-Kad{\i}k\"oy, \\ 34722 Istanbul, Turkey\\
$^*$ Physics Division, TOBB
University of Economics and Technology,  Ankara, Turkey\\
 and\\Institute of Physics, National Academy of Sciences,  Baku, Azerbaijan\\
$^1$bashiry@ciu.edu.tr\\
$^2$kazizi@dogus.edu.tr\\
$^3$ssultansoy@etu.edu.tr}

\begin{abstract}
The heavy  fourth generation of quarks that  have sufficiently small mixing with the three known  SM  families  form hadrons. In the present work, we calculate the masses and decay constants
of mesons containing  either both quarks from the fourth generation or one from  fourth family and the other from known third family SM quarks in the framework of the QCD
 sum rules. In the calculations, we take into account  two gluon condensate diagrams as nonperturbative contributions. The obtained  results   reduce to the known masses
and decay constants of the
$\bar b b$ and $\bar c c$  quarkonia when the fourth family quark is replaced by the bottom or charm quark.

\end{abstract}
\pacs{ 11.55.Hx,  12.60.-i}

\maketitle

%%%%%%%%%%%%%%%%%%%%%%%%%%%%%%%%%%%%%%%%%%%%%%%%%%%%%%%%%%%%%%%%%%%%
%\section{Introduction}
%%%%%%%%%%%%%%%%%%%%%%%%%%%%%%%%%%%%%%%%%%%%%%%%%%%%%%%%%%%%%%%%%%%%
\section{Introduction}
In the standard model (SM), we have three generation of quarks  experimentally observed. Among these quarks, the top ($t$) quark does not form bound states (hadrons) as a consequence of the high value of its mass. The top quark
immediately decays to the bottom quark giving a $W$ boson and this transition has full strength. The number of quark and lepton generations is one of the mysteries of  nature
 and can not be addressed by the SM. There are flavor democracy arguments that predict the existence of the  fourth generation of quarks \cite{ek1,ek2,Celikel}. It is expected that the masses of the fourth
 generation quarks are in the interval $(300-700)~GeV$ \cite{Sultansoy}, in which the upper limit coincides with the one obtained from partial-wave unitarity at high energies \cite{Chanowitz1}.  Within the
 flavor democracy approach, the Dirac masses of the fourth family fermions are almost equal, whereas masses of the first three family fermions as well as the CKM and PMNS mixings are obtained via small
 violations of democracy \cite{Datta,Atag}. For the recent status of the SM with fourth generation (SM$_4$), see e.g. \cite{ek3,ek4,ek5} and references therein.

Although the masses of fourth generation quarks are larger than the top quark mass (the last analysis of the Tevatron data implies $m_{d_4} > 372~ GeV$ \cite{Aaltonen} and $m_{u_4} > 358 ~GeV$ \cite{Convay}),
 they can form bound states as a result of the smallness of the mixing between these quarks and ordinary SM quarks \cite{atlas,ek6,ek7,Sultansoy2,ek8,ek9,Ishiwata}. As the mass difference between these
 two quarks is small, we will refer to   both members of the fourth family by $u_4$.  The condition for formation of new hadrons containing ultra-heavy quarks ($Q$)
is given by \cite{Bigi}:
\begin{equation}
\left| V_{Qq}\right|\leq \left(\frac{100~GeV}{m_{Q}}\right)
^{3/2}.
\end{equation}

For t-quark with $m_t = 172 ~GeV$, Eq. (1) leads to $V_{tq} < 0.44$, whereas the  single top production at the Tevatron gives $V_{tb} > 0.74$ \cite{ek10}.
When the fourth family quarks  have  sufficiently small mixing with the ordinary quarks, the hadrons made up from these quarks can live longer enough, and the bound state $\bar{u_4} u_4$ decays through its annihilation
and not via $u_4$ decays to a lower family quark plus a $W$ boson \cite{Ishiwata}. Concerning the flavor democracy approach, this situation is realized for parameterizations proposed
in \cite{Atag} and \cite{ek11}, whereas parameterization in \cite{Datta} predicts $V_{u_4q} \sim 0.2$ which does not allow formation of the fourth family quarkonia for $m_{u_4} > 300 GeV$.

Considering the above discussions, the production of
 such bound states if they exist will be possible at LHC. The conditions for observation of the
fourth SM family quarks at the LHC has been discussed in \cite{atlas,arik,ek12,ek13,ek14,ek15,ek16,ek17,ek18}. As there is a possibility to observe the bound states which consist of fourth family quarks at the LHC, it is reasonable to investigate their
 properties, theoretically and phenomenologically.

In the present work, we calculate the masses and decay constants
of the bound state mesons containing two  heavy quarks  either both  from the SM$_4$ or one from heavy fourth family and the other from ordinary heavy $b$ or $c$ quark. Here, we consider the ground state mesons
with different quantum numbers, namely scalars ($\bar{u}_4u_4$, $\bar{u}_4 b$ and $\bar{u}_4c$), pseudoscalars ($\bar{u}_4\gamma_5u_4$, $\bar{u}_4 \gamma_5b$ and $\bar{u}_4 \gamma_5c$), vectors
 ($\bar{u}_4\gamma_\mu u_4$, $\bar{u}_4 \gamma_\mu b$ and $\bar{u}_4 \gamma_\mu c$) and axial vector ($\bar{u}_4\gamma_\mu\gamma_5u_4$, $\bar{u}_4\gamma_\mu \gamma_5b$ and $\bar{u}_4 \gamma_\mu\gamma_5c$) mesons.
 These mesons, similar to the ordinary
hadrons, are formed in low energies very far from the asymptotic region. Therefore, to
calculate their hadronic parameters such as their masses and leptonic decay constants, we need to consult  some nonperturbative approaches. Among the nonperturbative methods, the QCD sum rules \cite{shifman},
 which is based on QCD Lagrangian and is free of model dependent parameters, is one of the most applicable and predictive approaches to hadron physics. This method has been successfully
 used to calculate the masses and decay constants of mesons both in vacuum and at finite temperature
(see for instance \cite{AIVainshtein,LJReinders,SNarison,
MJamin,AAPenin,Du,Kazem1,Kazem2,Kazem3,Kazem4}). Now, we extend the application of this method to calculate the masses and decay constants of the considered mesons containing fourth family quarkonia. The heavy quark
condensates are suppressed by the inverse powers of the heavy quark mass. Therefore, as the first nonperturbative contributions, we take into account  the two-gluon condensate diagrams.

 The outline of the paper is as follows.  In the
 next section, QCD sum rules for   masses and  decay constants of the considered bound states are obtained.
Section III encompasses our numerical analysis  on the masses and
 decay constants of the ground state ultra heavy scalar, pseudoscalar, vector and axial vector mesons  as well as our discussions.
%%%
%%%
\section{QCD sum rules for  masses and decay constants of the bound states (mesons) containing heavy fourth family quarks }
We start to this section considering  sufficient correlation functions responsible for calculation of the masses and decay constants of the bound states containing heavy fourth generation quarks in
the framework of  QCD sum rules. The two point correlation function corresponding to the scalar (S) and pseudoscalar (PS) cases is written as:
\begin{eqnarray}\label{correl.func.1}
\Pi^{S(PS)} =i\int d^{4}xe^{ip. x}{\langle} 0\mid{\cal T}\left ( J^{S(PS)} (x) \bar J^{S(PS)}(0)\right)\mid0{\rangle},
\end{eqnarray}
where  ${\cal T}$  is
the time ordering product and
$J^S(x)=\overline{u}_4(x)q(x)$ and $J^{PS}(x)=\overline{u}_4(x)\gamma_5q(x)$ are the interpolating currents of the
heavy scalar and pseudoscalar bound states, respectively. Here,  the $q$ can be either fourth family $u_4$ quark or ordinary heavy $b$ or $c$ quark. Similarly, the correlation function for the vector (V) and axial 
vector (AV)  is written as:
\begin{eqnarray}\label{correl.func.11}
\Pi^{V(AV)}_{\mu\nu} =i\int d^{4}xe^{ip. x}{\langle} 0\mid{\cal T}\left ( J_\mu^{V(AV)} (x) \bar J_\nu^{V(AV)}(0)\right)\mid0{\rangle},
\end{eqnarray}
where, the currents $J_\mu^{V}=\overline{u}_4(x)\gamma_\mu q(x)$ and  $J_\mu^{AV}=\overline{u}_4(x)\gamma_\mu\gamma_5 q(x)$ are responsible for creating the vector and axial vector quarkonia from  vacuum with the same
quantum numbers as the interpolating currents.

From the general philosophy of the QCD sum rules, we calculate the aforesaid correlation functions in two alternative ways. From the physical or phenomenological side, we calculate them in terms of hadronic parameters
such as masses and decay constants. In QCD or theoretical side, they are calculated in terms of QCD degrees of freedom such as quark masses and gluon condensates by the help of operator product expansion (OPE) in
deep Euclidean region. Equating these two representations of the correlation functions through dispersion relations, we acquire the QCD sum rules for the masses and decay constants. These sum rules relate the hadronic
parameters to the fundamental QCD parameters. To suppress  contribution of the higher states and continuum,  Borel transformation with respect to the momentum squared is applied to both sides of the correlation
 functions.

First, to calculate the phenomenological part, we  insert a complete
set of intermediate  states having the same quantum numbers as the interpolating  currents to the correlation functions. Performing the integral over $x$ and isolating the ground state, we obtain
\begin{eqnarray}\label{phen1}
\Pi^{S(PS)}=\frac{{\langle}0\mid J^{S(PS)}(0) \mid S(PS)\rangle \langle S (PS)\mid J^{S(PS)}(0)\mid
 0\rangle}{m_{S(PS)}^2-p^2}
&+& \cdots,
\end{eqnarray}
 where $\cdots$ represents  contributions of the  higher states and continuum and $m_{S(PS)}$ is  mass of the heavy scalar(pseudoscalar) meson.
From a similar manner, for the vector (axial vector) case, we obtain
\begin{eqnarray}\label{phen1}
\Pi^{V(AV)}_{\mu\nu}=\frac{{\langle}0\mid J^{V(AV)}_\mu(0) \mid V(AV)\rangle \langle V(AV)\mid J^{V(AV)}_\nu(0)\mid
 0\rangle}{m_{V(AV)}^2-p^2}
&+& \cdots,
\end{eqnarray}
To proceed, we need to know the matrix elements of the interpolating currents between the vacuum and mesonic states. These matrix elements are parametrized in terms of  leptonic decay constants as:
\begin{eqnarray}\label{lep}
\langle 0 \mid J(0)\mid S\rangle&=&f_{S}  m_{S},\nonumber\\
\langle 0 \mid J(0)\mid PS\rangle&=&f_{PS}  \frac{m^2_{PS}}{m_{u4}+m_q},\nonumber\\
\langle 0 \mid J(0)\mid V(AV)\rangle&=&f_{V(AV)}   m_{V(AV)} \varepsilon_\mu,
\end{eqnarray}
where $f_i$ are the leptonic decay constants of the considered bound state mesons.
Using  summation over  polarization vectors in the $V(AV)$ case as
\begin{eqnarray}\label{polvec}
\epsilon_\mu\epsilon^{*}_\nu&=&-g_{\mu\nu}+\frac{p_{\mu}p_{\nu}}{m_{V(AV)}^2},\nonumber\\
\end{eqnarray}
we get, the final expressions of the physical sides of the correlation functions as:
\begin{eqnarray}\label{phen2}
\Pi^{S}&=&\frac{f_{S}^2 m_{S}^2}{m_{S}^2-p^2} + \cdots\nonumber\\
\Pi^{PS}&=&\frac{f_{PS}^2 (\frac{m_{PS}^2}{m_{u4}+m_q})^2}{m_{PS}^2-p^2} + \cdots\nonumber\\
\Pi^{V(AV)}_{\mu\nu}&=&\frac{f_{V(AV)}^2 m_{V(AV)}^2}{m_{V(AV)}^2-p^2}\left[-g_{\mu\nu}+\frac{p_{\mu}p_{\nu}}{m_{V(AV)}^2}\right] + \cdots,\nonumber\\
\end{eqnarray}
where to calculate the mass and decay constant in the $V(AV)$ channel, we choose the structure $g_{\mu\nu}$.

\begin{figure}[h!]
\begin{center}
\includegraphics[width=12cm]{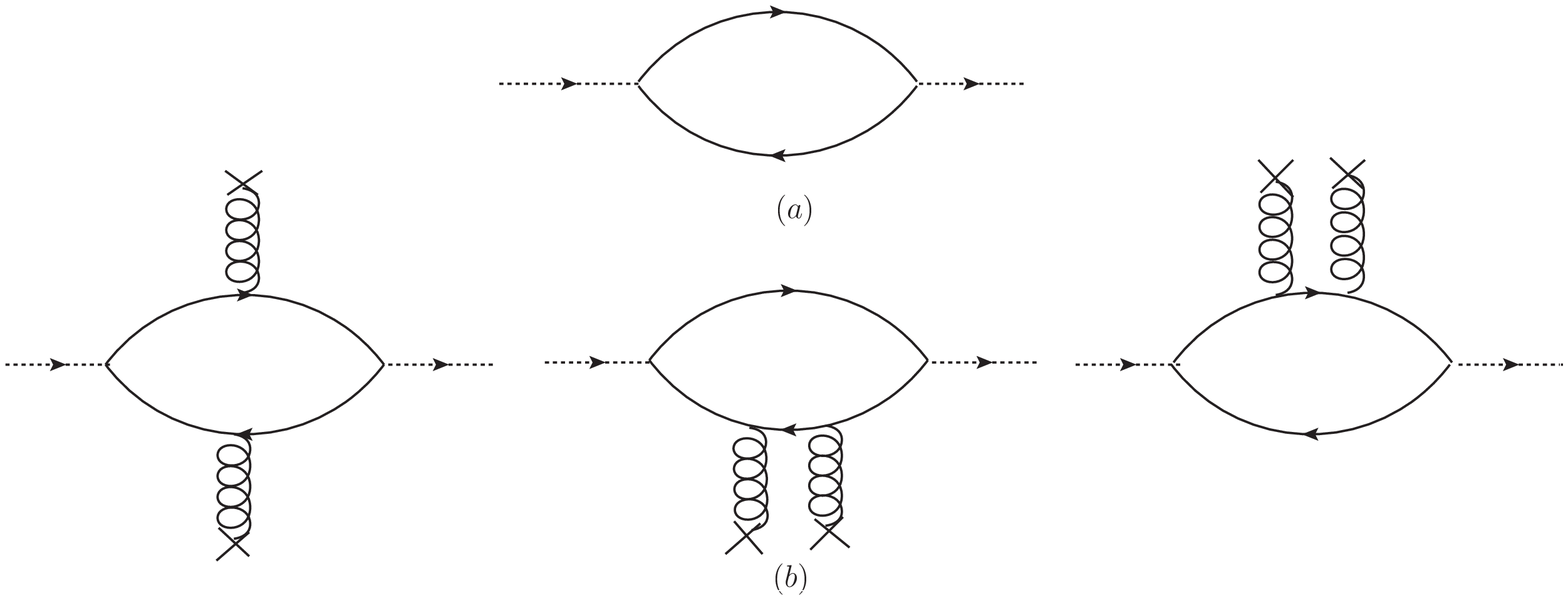}
\end{center}
\caption{ (a): Bare loop diagram
 (b): Diagrams corresponding to  gluon condensates.} \label{fig1}
\end{figure}

 In QCD side, the correlation functions are  calculated in deep Euclidean region, $p^2\ll-\Lambda_{QCD}^2$ via OPE, where  short or perturbative and long distance or non-perturbative effects are
 separated. For each correlation function in $S(PS)$ case and coefficient of the selected structure in $V(AV)$ channel, we write
\begin{eqnarray}\label{correl.func.QCD1}
\Pi^{QCD} =\Pi_{pert}+\Pi_{nonpert}.
\end{eqnarray}
The short distance contribution  (bare loop diagram  in figure
(\ref{fig1}) part (a)) in each case is  calculated using the perturbation theory,
 whereas the long distance   contributions (diagrams shown in figure
(\ref{fig1}) part  (b) ) are parameterized in terms of   gluon condensates. To proceed, we write the perturbative part in terms of a dispersion
integral,
\begin{eqnarray}\label{correl.func.QCD1}
\Pi^{QCD} =\int \frac{ds \rho(s)}{s-p^2}+\Pi_{nonpert},
\end{eqnarray}
where, $\rho(s)$  is called the spectral density. To calculate the spectral density, we calculate the Feynman amplitude of the bare loop diagram by the help of  Cutkosky
 rules, where the quark propagators are replaced by Dirac delta function, i.e.,  $\frac{1}{p^2-m^2}\rightarrow (- 2\pi i) \delta(p^2-m^2)$.
As a result, the spectral
density is obtained as follows:
\begin{equation}
\rho(s)=\frac{3s}{8\pi^2}(1-\frac{(m_1\pm m_2)^2}{s})\sqrt{1-2 \frac{m_1^2+m_2^2}{s}+\frac{(m_1^2-m_2^2)^2}{s^2}}
\end{equation}
where $+$ sign in $(m_1\pm m_2)$ is chosen for scalar and axial vector cases  and $-$ sign is for psoduscalar and vector channels. Here, $m_1=m_{u_4}$ and $m_2$ is either $m_{u_4}$ or $m_{c(b)}$.

To obtain the non-perturbative part, we calculate the gluon condensate diagrams represented in part (b) of
figure (\ref{fig1}). For this aim, we use Fock-Schwinger
gauge,  $x^{\mu}A^{a}_{\mu}(x)=0$. In momentum space, the  vacuum gluon field is expressed as:
\begin{eqnarray}\label{Amu}
A^{a}_{\mu}(k')=-\frac{i}{2}(2 \pi)^4 G^{a}_{\rho
\mu}(0)\frac{\partial} {\partial k'_{\rho}}\delta^{(4)}(k'),
\end{eqnarray}
where $k'$ is the gluon momentum. 
In the calculations, we also use the  quark-gluon-quark vertex as:
\begin{eqnarray}\label{qgqver}
\Gamma_{ij\mu}^a=ig\gamma_\mu
\left(\frac{\lambda^{a}}{2}\right)_{ij},
\end{eqnarray}
After straightforward but lengthy
calculations, the non-perturbative part for each channel in momentum space is obtained as:

\begin{eqnarray}
\Pi_{nonpert}^i=\int^1_0\langle \alpha_s
G^2\rangle\frac{\Theta^i+\Theta^i(m_1\leftrightarrow m_2)}{96\pi(m_2^2 + m_1^2 x - m_2^2 x - p^2 x + p^2 x^2)^4}dx
\end{eqnarray}
where $\Theta^i(m_1\leftrightarrow m_2)$ means that in $\Theta^i$, we exchange $m_1$ and $m_2$. The explicit expressions for $\Theta^i$ are given as:
  \begin{eqnarray}
\Theta^S &&=\frac{1}{2} x^2 \Bigg\{3 m_1^4 x (m_2^2 (x (17-2
   x (2 x (9 x-26)+47))+8)\nonumber\\ &&+p^2 x (x (27 x-25)-7)
   (x-1)^2)+2 m_2 m_1^3 (m_2^2 (x (x (x
   (21 x-58)+39)\nonumber\\ &&+12)-15) -p^2 (x-1) x (x (x (7
   x-13)-3)+12)) \nonumber\\ &&+m_1^2 (-m_2^2 p^2 (x-1)
  x (x (x (2 x (81 x-242)+455)-96)-33) \nonumber\\ &&+m_2^4 (x (x
   (x (3 x (36 x-145)+652)-414)+72)+15)+3 p^4
   (x-1)^3\nonumber\\ && x^2 (24 x^2-22 x-5)) -m_2
   m_1 (x-1) (-m_2^2 p^2 (x^2-2) (x
   (14 x-27)+15)\nonumber\\ &&+m_2^4 (3 x-5) (x (7 x-12)+6)+p^4
   (x-1) x (x (2 x (7 x-13)+3)+12))\nonumber\\ &&+(x-1)
   (-m_2^2 p^4 (x-1) x (2 x (x (2 x (18
   x-55)+109)-30)-9)\nonumber\\ &&+m_2^4 p^2 (x (x (x (x (81
   x-328)+490)-299)+42)+15)\nonumber\\ &&-m_2^6 (2 x-3) (x (6 x (3
   x-8)+47)-15)+3 p^6 (x-1)^3 x^2 (6 (x-1)
   x-1))\nonumber\\ &&+9 m_1^6 (x-1)^2 x^2 (4 x+1)+3 m_2
   m_1^5 x (x ((8-7 x) x+2)-4)\Bigg\},\nonumber\\
%\end{eqnarray}
%
%\begin{eqnarray}
\Theta^{PS} &&=-\frac{1}{2} x^2 \Bigg\{-3 m_1^4 x (m_2^2
   (36 x^4-104 x^3+94 x^2-17 x-8)\nonumber\\ &&-p^2
   (x-1)^2 x (27 x^2-25 x-7))-2 m_2
   m_1^3 (m_2^2 (21 x^4-58 x^3+39 x^2+12
   x-15)\nonumber\\ &&+p^2 x (-7 x^4+20 x^3-10 x^2-15
   x+12))+m_2 m_1 (x-1) (m_2^2 p^2
   (-14 x^4\nonumber\\ &&+27 x^3+13 x^2-54 x+30)+m_2^4
   (21 x^3-71 x^2+78 x-30)\nonumber\\ &&+p^4 x (14
   x^4-40 x^3+29 x^2+9 x-12))+m_1^2
   (-m_2^2 p^2 x (162 x^5-646 x^4+939
   x^3\nonumber\\ &&-551 x^2+63 x+33)+m_2^4 (108
   x^5-435 x^4+652 x^3-414 x^2+72 x+15)\nonumber\\ &&+3 p^4
   (x-1)^3 x^2 (24 x^2-22
   x-5))+(x-1) (-m_2^2 p^4 x
   (72 x^5-292 x^4 +438 x^3\nonumber\\ &&-278 x^2+51
   x+9)+m_2^4 p^2 (81 x^5-328 x^4+490
   x^3-299 x^2+42 x+15)\nonumber\\ &&+m_2^6 (-36
   x^4+150 x^3-238 x^2+171 x-45)+3 p^6 (x-1)^3
   x^2 (6 x^2-6 x-1))\nonumber\\ &&+9 m_1^6
   (x-1)^2 x^2 (4 x+1)+3 m_2 m_1^5 x (7 x^3-8
   x^2-2 x+4)\Bigg\},\nonumber\\
%\end{eqnarray}
%%
%\begin{eqnarray}
\Theta^{V} &&=-\frac{1}{2} (x-1)^2 \Bigg\{m_1^4 x^2 (m_2^2 (2 x (1-18 (x-1)
   x)+3)\nonumber\\ &&+p^2 (x (27 x-25)-7) x^2)+2 m_2 m_1^3 (x-1)^2 x
   (m_2^2 (3 x-4)\nonumber\\ &&-p^2 (x-3) x)-m_2 m_1 (x-1)^2
   (m_2^2 p^2 x ((7-2 x) x-8)+m_2^4 (x-1) (3 x-5)\nonumber\\ &&+p^4 x^2 (2
   (x-1) x+3))+m_1^2 (x-1) x (m_2^2 p^2 x (x
   (-54 x^2+56 x+5)+4)\nonumber\\ &&+m_2^4 (9 (x-1) x (4
   x-1)-8)+p^4 x^3 (24 x^2-22 x-5))+(x-1)^2\nonumber\\ &&
   (m_2^2 p^4 x^2 (4 (7-6 x) x^2+1)+m_2^4 p^2 x
   (x^2 (27 x-31)-3)\nonumber\\ &&+m_2^6 (5-2 x (6 x^2-9
   x+4))+p^6 x^4 (6 (x-1) x-1))\nonumber\\ &&+3 m_1^6 x^4 (4
   x+1)-3 m_2 m_1^5 (x-1)^2 x^2\Bigg\},\nonumber\\
%\end{eqnarray}
%and
%\begin{eqnarray}
\Theta^{AV}&&=-\frac{1}{2} x^2 \Bigg\{2 m_2 m_1^3 x^3
   (m_2^2 (4-3 x)+p^2
   (x^2+x-2))\nonumber\\ &&+m_1^4 x
   (m_2^2 (x (17-2 x (18 (x-3)
   x+47))+8)+p^2 x (x (27 x-25)-7)
   (x-1)^2)\nonumber\\ &&+m_1^2 (-m_2^2 p^2 (x-1) x
   (x (x (2 x (27 x-82)+149)-32)-11)\nonumber\\ &&+m_2^4 (3 x
   (x (x (3 x (4 x-17)+76)-46)+8)+5)+p^4 (x-1)^3
   x^2 (24 x^2-22 x-5))\nonumber\\ &&+m_2 m_1
   (x-1) x^2 (m_2^2 p^2 (7-x (2 x+3))+m_2^4
   (3 x-5)\nonumber\\ &&+p^4 (x-1) (2 (x-1) x+3))+(x-1)
   (-m_2^2 p^4 (x-1) x (2 x (x (2 x (6
   x-19)\nonumber\\ &&+37)-10)-3)+m_2^4 p^2 (x (x (x (x (27
   x-112)+162)-97)+14)+5)\nonumber\\ &&+m_2^6 (x (57-2 x (3 x
   (2 x-9)+43))-15)+p^6 (x-1)^3 x^2 (6 (x-1)
   x-1))\nonumber\\ &&+3 m_2 m_1^5 x^4+3 m_1^6 (x-1)^2
   x^2 (4 x+1)\Bigg\}.
\end{eqnarray}

 The next step is to match the phenomenological and QCD sides of the correlation
functions to get  sum rules  for the masses and decay constants  of  the bound states. To suppress  contribution of the higher
states and continuum,  Borel transformation over  $p^2$ as
well as continuum subtraction are performed. As a result of this procedure, we obtain the following sum rules:
\begin{eqnarray}\label{lepsum}
m_{S(V)(AV)}^2f_{S(V)(AV)}^2e^{\frac{-m_{S(V)(AV)}^2}{M^2}}&=&
\int_{(m_1+m_2)^2}^{s_{0}}
ds~\rho^{S(V)(AV)}(s)~e^{-\frac{s}{M^{2}}}+\hat{B}\Pi^{S(V)(AV)}_{nonpert},\nonumber\\
\frac{m_{PS}^4f_{PS}^2}{(m_{u_4}+m_q)^2}e^{\frac{-m_{PS}^2}{M^2}}&=&
\int_{(m_1+m_2)^2}^{s_{0}}
ds~\rho^{PS}(s)~e^{-\frac{s}{M^{2}}}+\hat{B}\Pi^{PS}_{nonpert},
\end{eqnarray}
where $M^2$ is the Borel mass parameter and $s_{0}$ is the  continuum threshold. The sum rules for the masses are obtained
applying derivative with respect to $-\frac{1}{M^2}$  to the both
sides of the above  sum rules  and dividing by themselves. i.e.,
\begin{eqnarray}\label{mass2}
m_{S(PS)(V)(AV)}^2=\frac{-\frac{d}{d(\frac{1}{M^2})}\left[\int_{(m_1+m_2)^2}^{s_{0}}
ds~\rho^{S(PS)(V)(AV)}(s)~e^{-\frac{s}{M^{2}}}+\hat{B}\Pi^{S(PS)(V)(AV)}_{nonpert}\right]}{\int_{(m_1+m_2)^2}^{s_{0}}
ds~\rho^{S(PS)(V)(AV)}(s)~e^{-\frac{s}{M^{2}}}+\hat{B}\Pi^{S(PS)(V)(AV)}_{nonpert}},
\end{eqnarray}
where
\begin{eqnarray}
\hat{B}\Pi^i_{nonpert}=\int^1_0 e^{\frac{m_2^2+x(m_1^2-m_2^2)}{M^2x(x-1)}}\frac{
 \Delta^i+ \Delta^i(m_1\leftrightarrow m_2)}{\pi 96 M^6(x-1)^4x^3}\langle \alpha_s G^2\rangle dx,
\end{eqnarray}
and
\begin{eqnarray}
\Delta^S&&=-m_2 m_1^3 (x-1) x^2 (m_2^2 (14 x^2-29
   x+14)\nonumber\\ &&+2 M^2 x (7 x^2-13
   x+6))+m_1^4 (x-1) x^3 (m_2^2
   (9 x^2-14 x+6)\nonumber\\ &&+3 M^2 x (3 x^2-4
   x+1))+m_2 m_1 (x-1) (m_2^2 M^2 x\nonumber\\ &&
   (14 x^4-53 x^3+71 x^2-36 x+6)+m_2^4
   (7 x^4-28 x^3+40 x^2-25 x+6)\nonumber\\ &&+2 M^4 x^2
   (14 x^4-40 x^3+29 x^2+9
   x-12))+m_1^2 x (m_2^2 M^2 x\nonumber\\ &&
   (-18 x^5+70 x^4-105 x^3+77 x^2-27
   x+3)+m_2^4 (-9 x^5+37 x^4\nonumber\\ &&-61 x^3+52
   x^2-21 x+3)-12 M^4 x^2 (3 x+1)
   (x-1)^4)-(x-1) \nonumber\\ &&(-2 m_2^2 M^4 x^3
   (18 x^4-76 x^3+123 x^2-89 x+24)\nonumber\\ &&+m_2^4
   M^2 x (-9 x^5+40 x^4-71 x^3+68 x^2-33
   x+6)+m_2^6 (-3 x^5+14 x^4\nonumber\\ &&-27 x^3+29
   x^2-15 x+3)+6 M^6 (x-1)^3 x^3 (6 x^2-6
   x-1))\nonumber\\ &&-3 m_1^6 (x-1) x^5+m_2 m_1^5 x^3
   (7 x^2-8 x+1),\nonumber\\
%\end{eqnarray}
%\begin{eqnarray}
\Delta^{PS} &&=-m_2 m_1^3 (x-1) x^2 (m_2^2 (14 x^2-29
   x+14)\nonumber\\ &&+2 M^2 x (7 x^2-13
   x+6))-m_1^4 (x-1) x^3 (m_2^2
   (9 x^2-14 x+6)\nonumber\\ &&+3 M^2 x (3 x^2-4
   x+1))+m_2 m_1 (x-1) (m_2^2 M^2 x
   (14 x^4-53 x^3+71 x^2-36 x+6)\nonumber\\ &&+m_2^4
   (7 x^4-28 x^3+40 x^2-25 x+6)+2 M^4 x^2
   (14 x^4-40 x^3+29 x^2+9
   x-12))\nonumber\\ &&+m_1^2 x (m_2^2 M^2 x
   (18 x^5-70 x^4+105 x^3-77 x^2+27
   x-3)+m_2^4 (9 x^5-37 x^4\nonumber\\ &&+61 x^3-52
   x^2+21 x-3)+12 M^4 x^2 (3 x+1)
   (x-1)^4)\nonumber\\ &&+(x-1) (-2 m_2^2 M^4 x^3
   (18 x^4-76 x^3+123 x^2-89 x+24)\nonumber\\ &&+m_2^4
   M^2 x (-9 x^5+40 x^4-71 x^3+68 x^2-33
   x+6)\nonumber\\ &&+m_2^6 (-3 x^5+14 x^4-27 x^3+29
   x^2-15 x+3)\nonumber\\ &&+6 M^6 (x-1)^3 x^3 (6 x^2-6
   x-1))+3 m_1^6 (x-1) x^5+m_2 m_1^5 x^3
   (7 x^2-8 x+1),\nonumber\\
%\end{eqnarray}
%\begin{eqnarray}
\Delta^{V} &&=m_2 m_1^3 (x-1)^2 x^2 (m_2^2 (2 x-1)+2 M^2 x
   (x+2))\nonumber\\ &&-m_1^4 (x-1) x^3 (m_2^2 (3 x^2-3
   x+1)+M^2 (3 x-1) x^2)-m_2 m_1 (x-1)^3 x\nonumber\\ &&
   (m_2^2 M^2 (2 x^2+3 x-2)+m_2^4 (x-1)+2 M^4 x
   (2 x^2-2 x+3))\nonumber\\ &&+m_1^2 (x-1)^2 x (m_2^2
   M^2 x (6 x^3-8 x^2+x+2)+m_2^4 (3 x^3-6 x^2+4
   x-1)\nonumber\\ &&+4 M^4 x^3 (3 x^2-2 x-1))+(x-1)^3
   (2 m_2^2 M^4 x^2 (-6 x^3+10 x^2-3 x+1)\nonumber\\ &&-m_2^4
   M^2 x (3 x^3-7 x^2+3 x+1)-m_2^6 (x-1)^3+2 M^6 x^4
   (6 x^2-6 x-1))\nonumber\\ &&+m_1^6 x^6-m_2 m_1^5 (x-1) x^4,\nonumber\\
%\end{eqnarray}
%\begin{eqnarray}
\Delta^{AV}&&=-m_2 m_1^3 (x-1) x^2 (m_2^2 (2 x^2-5
   x+2)\\ &&+2 M^2 x (x^2-4
   x+3))-m_1^4 (x-1) x^3 (m_2^2
   (3 x^2-6 x+2)\nonumber\\ &&+M^2 x (3 x^2-4
   x+1))+m_2 m_1 (x-1) x (m_2^2
   M^2 x (2 x^3-11 x^2+17 x-6)\nonumber\\ &&+m_2^4
   (x^3-4 x^2+4 x-1)+2 M^4 x^2
   (2 x^3-4 x^2+5 x-3))+m_1^2 x
   (m_2^2 M^2 x (6 x^5\nonumber\\ &&-26 x^4+43
   x^3-31 x^2+9 x-1)+m_2^4 (3 x^5-15
   x^4+27 x^3-20 x^2+7 x-1)\nonumber\\ &&+4 M^4 x^2 (3
   x+1) (x-1)^4)+(x-1) (-2 m_2^2 M^4
   x^3 (6 x^4-28 x^3+45 x^2-31
   x+8)\nonumber\\ &&+m_2^4 M^2 x (-3 x^5+16 x^4-33
   x^3+28 x^2-11 x+2)-m_2^6 (x^5-6
   x^4+13 x^3-11 x^2\nonumber\\ &&+5 x-1)+2 M^6 (x-1)^3
   x^3 (6 x^2-6 x-1))+m_1^6
   (x-1) x^5+m_2 m_1^5 (x-1)^2 x^3.\nonumber
\end{eqnarray}

\section{Numerical Results}
To obtain numerical values for the masses and decay constants of the considered bound states containing heavy fourth family from the obtained QCD sum rules,  we take the mass of the $u_4$ in the
interval $m_{u_4}=(450-550)~GeV$, $m_b=4.8~GeV$, $m_c=1.3~GeV$ and ${\langle}0\mid \frac{1}{\pi}\alpha_s G^2
\mid 0 {\rangle}=0.012~GeV^4$. The sum rules  for the masses
and decay constants   contain also two auxiliary parameters, namely Borel mass parameter $M^2$ and continuum
threshold $s_0$.  The standard
criteria in QCD sum rules is that the physical quantities should be
independent of the  auxiliary parameters. Therefor, we should look for working regions of these parameters such that our results are approximately insensitive to their variations.
 The
working region for the Borel mass parameter is determined demanding
that not only  the higher states and continuum contributions are
suppressed but  contributions of the highest order operators
should also be small, i.e., the sum rules for the masses and decay constants
should converge. As a result of the above procedure, the
working region for the Borel parameter is found to be $ 500~ GeV^2
\leq M^2 \leq 900~ GeV^2 $ for $\bar{u}_4b$ and $\bar{u}_4c$, and $ 1200~ GeV^2 \leq M^2
\leq 2000~ GeV^2 $ for $\bar{u}_4u_4$ heavy SM$_4$ mesons.
The continuum
threshold $s_{0}$ is not completely arbitrary  but correlated to
the energy of the first exited state with the same quantum number as the interpolating current. We have no information about the energy of the first excitation of the bound states 
containing fourth family quarks.
 Hence, 
the only way to determine the working region   is to choose a region such that not only  the results depend weakly on this parameter   
but the dependence of the physical observables on the Borel parameter
$M^2$ is also minimal. Our numerical calculations lead to the interval
$(m_1+m_2+3.3)^2~GeV^2\leq s_0\leq(m_1+m_2+3.7)^2~GeV^2$ for the continuum threshold.

\begin{figure}[h!]
\begin{center}
\includegraphics[width=10cm]{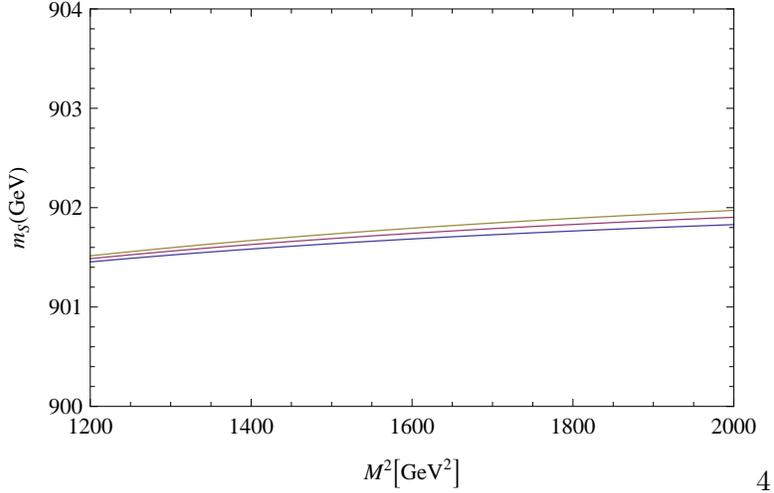}4
\end{center}
\caption{Dependence of mass of the scalar $\bar{u_4}u_4$
  on the Borel parameter, $M^{2}$  at three fixed
values of the continuum threshold. The upper, middle and lower lines belong to the values $s_0=(m_1+m_2+3.7)^2~GeV^2$, $s_0=(m_1+m_2+3.5)^2~GeV^2$ and $s_0=(m_1+m_2+3.3)^2~GeV^2$, respectively.} \label{fig2}
\end{figure}
%%%%%%%%%%%%%%%%%%%%%%%%%%%%%%%%%%%%%%%%%%%%%%%%%%%%%%%%%%%%%%%%%%%%%%%%%%%%%%%%%%%
%%%%%%%%%%%%%%%%%%%%%%%%%%%%%%%%%%%%%%%%%%%%%%%%%%%%%%%%%%%%%%%%%%%%%%%%%%%%%
\begin{figure}[h!]
\begin{center}
\includegraphics[width=10cm]{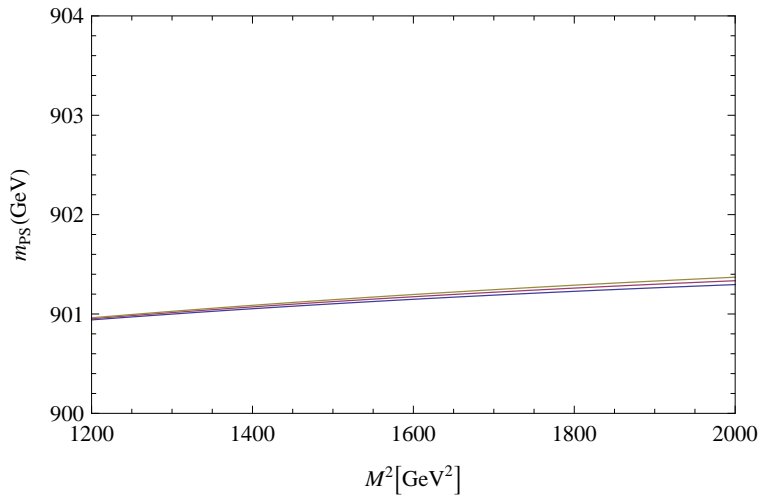}
\end{center}
\caption{The same as Fig. \ref{fig2} but for pseudoscaler $\bar{u_4}\gamma_5u_4$.} \label{fig3}
\end{figure}

%%%%%%%%%%%%%%%%%%%%%%%%%%%%%%%%%%%%%%%%%%%%%%%%%%%%%%%%%%4%%%%%%%%%%%%%%%%%%%%
\begin{figure}[h!]
\begin{center}
\includegraphics[width=10cm]{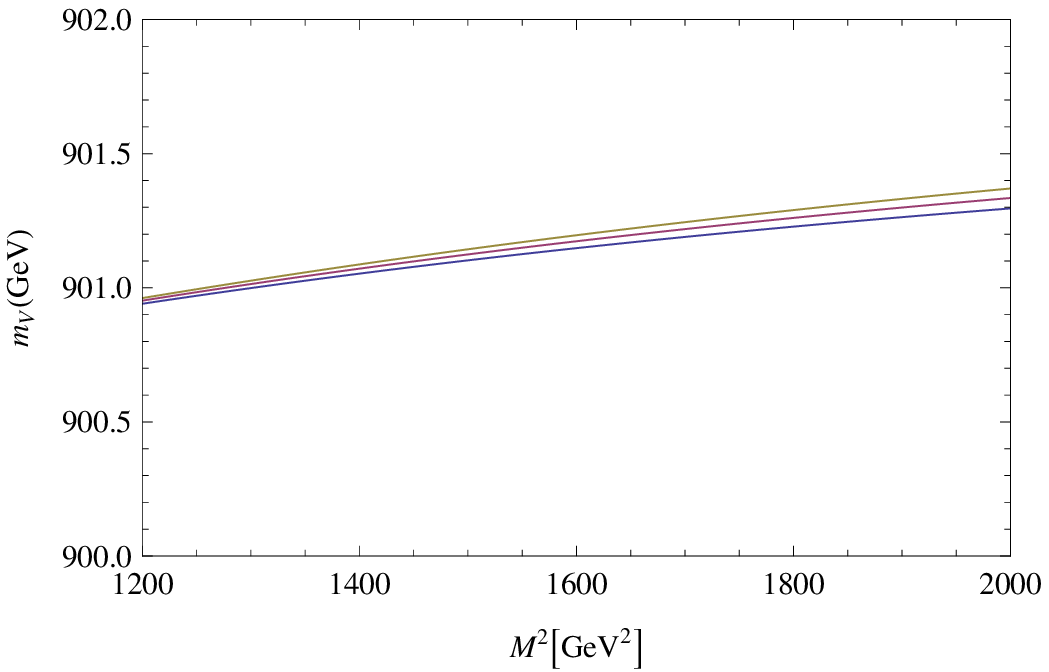}
\end{center}
\caption{The same as Fig. \ref{fig2} but for vector $\bar{u_4}\gamma_\mu u_4$.} \label{fig4}
\end{figure}

%%%%%%%%%%%%%%%%%%%%%%%%%%%%%%%%%%%%%%%%%%%%%%%%%%%%%%%%%%%%%%%%%%%%%%%%%%%%%%%%%%%
%%%%%%%%%%%%%%%%%%%%%%%%%%%%%%%%%%%%%%%%%%%%%%%%%%%%%%%%%%%%%%%%%%%%%%%%%%%%%%%%%%%
\begin{figure}[h!]
\begin{center}
\includegraphics[width=10cm]{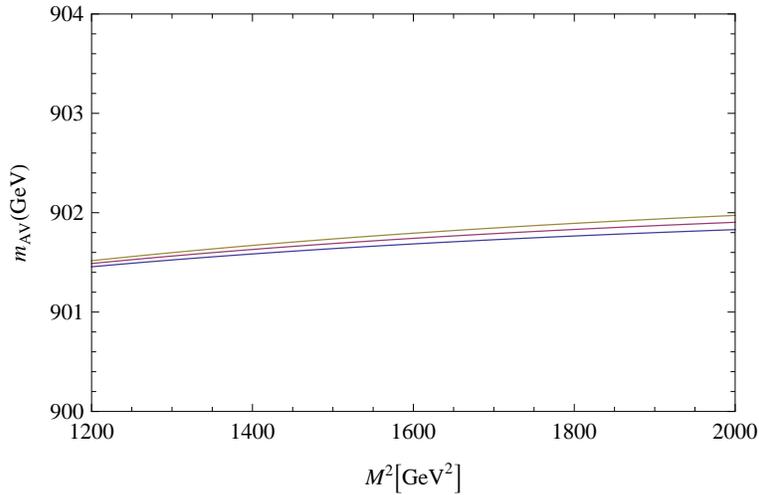}
\end{center}
\caption{The same as Fig. \ref{fig2} but for axial vector $\bar{u_4}\gamma_5 \gamma_\mu u_4$.} \label{fig5}
\end{figure}

%%%%%%%%%%%%%%%%%%%%%%%%%%%%%%%%%%%%%%%%%%%%%%%%%%%%%%%%%%%%%%%%%%%%%%%%%%%%%%%%%%%
%
\begin{figure}[h!]
\begin{center}
\includegraphics[width=10cm]{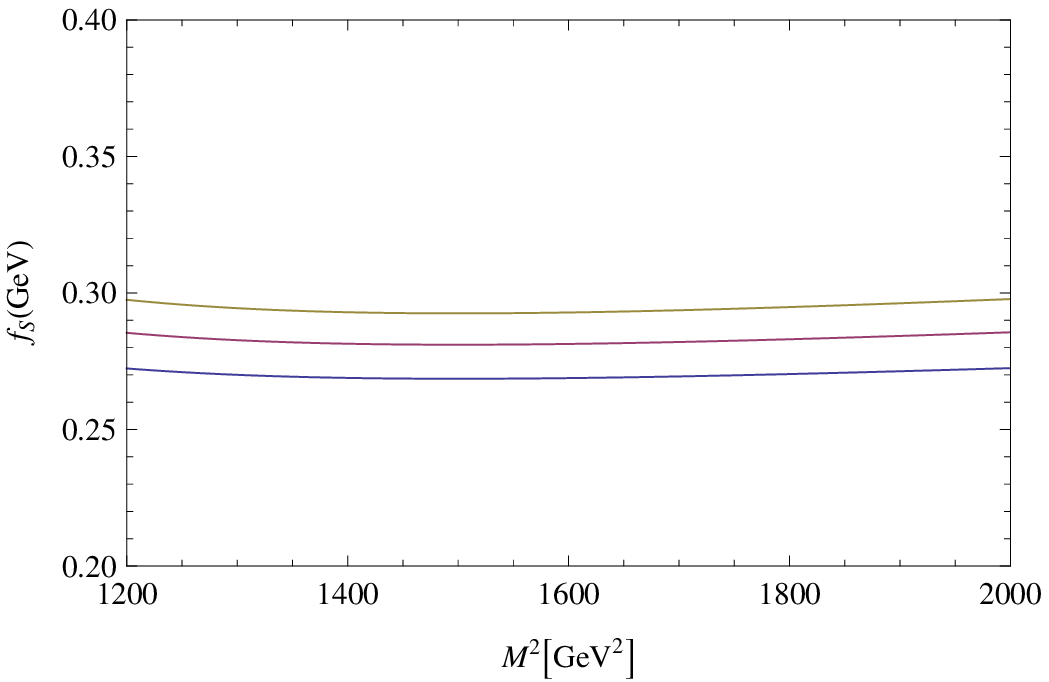}
\end{center}
\caption{Dependence of the decay constant of the scalar $\bar{u_4}u_4$
  on the Borel parameter, $M^{2}$  at three fixed
values of the continuum threshold. The upper, middle and lower lines belong to the values $s_0=(m_1+m_2+3.7)^2~GeV^2$, $s_0=(m_1+m_2+3.5)^2~GeV^2$ and $s_0=(m_1+m_2+3.3)^2~GeV^2$, respectively.} \label{fig6}
\end{figure}
%%%%%%%%%%%%%%%%%%%%%%%%%%%%%%%%%%%%%%%%%%%%%%%%%%%%%%%%%4%%%%%%%%%%%%%%%%%%%%%%%%%%%
\begin{figure}[h!]
\begin{center}
\includegraphics[width=10cm]{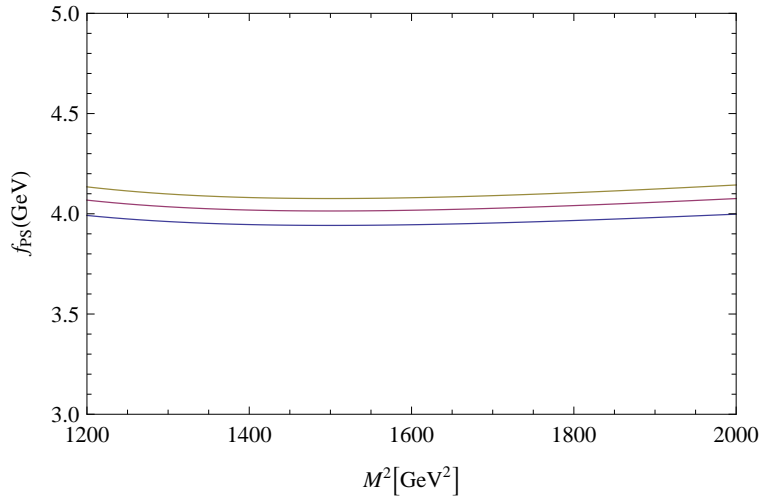}
\end{center}
\caption{The same as Fig. \ref{fig6} but for the decay constant of pseudoscalar $\bar{u_4}\gamma_5u_4$.} \label{fig7}
\end{figure}
%%%%%%%%%%%%%%%%%%%%%%%%%%%%%%%%%%%%%%%%%%%%%%%%%%%%%%%%%%%%%%%%%%%%%%%%%%%%%%%%%%%%
\begin{figure}[h!]
\begin{center}
\includegraphics[width=10cm]{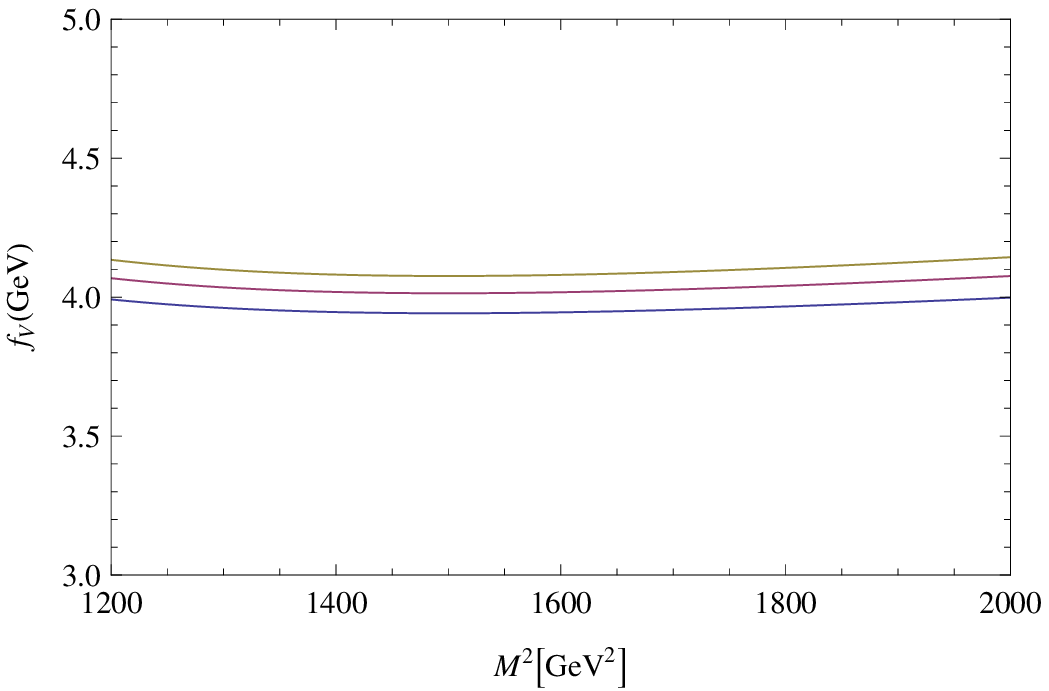}
\end{center}
\caption{The same as Fig. \ref{fig6} but for the decay constant  of vector $\bar{u_4}\gamma_\mu u_4$.} \label{fig8}
\end{figure}
%%%%%%%%%%%%%%%%%%%%%%%%%%%%%%%%%%%%%%%%%%%%%%%%%%%%%%%%%%%%%%%%%%%%%%%%%%%%%%%%%%%%
\begin{figure}[h!]4
\begin{center}
\includegraphics[width=10cm]{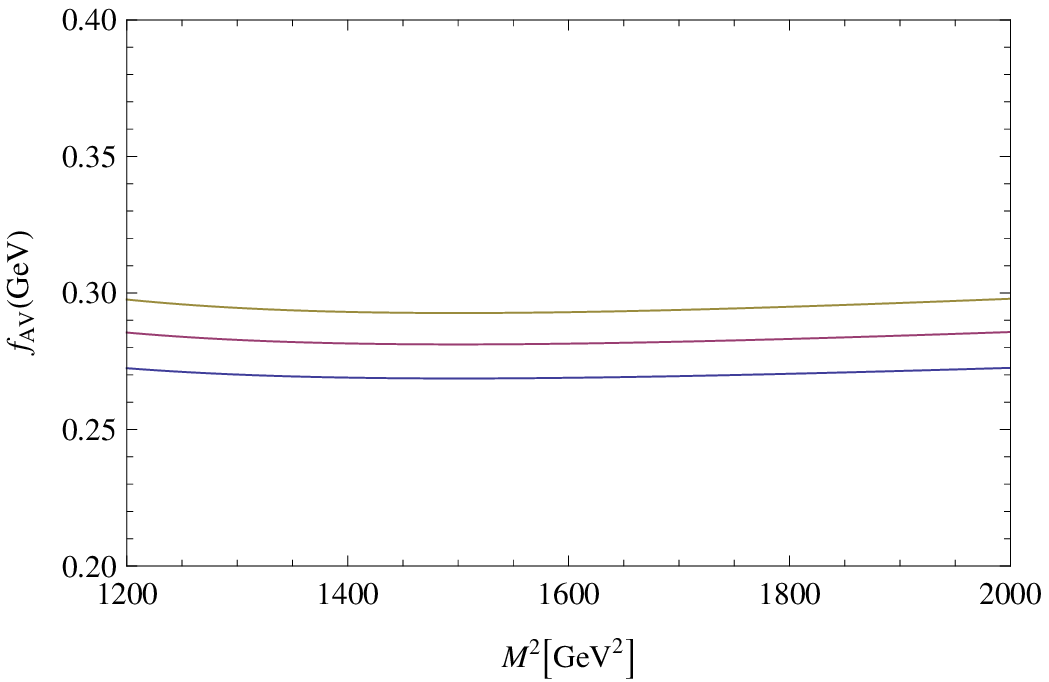}
\end{center}
\caption{The same as Fig. \ref{fig6} but for the decay constant of axial vector $\bar{u_4}\gamma_5\gamma_\mu u_4$.} \label{fig9}
\end{figure}
%%%%%%%%%%%%%%%%%%%%%%%%%%%%%%%%%%%%%%%%%%%%%%%%%%%%%%%%%%%%%%%%%%%%%%%%%%%%%%%%%%%%%%%%%

As an example, let us consider the case of the bound state $\bar{u_4}u_4$. The dependence of the masses of scalar $\bar{u_4}u_4$, pseudoscalar $\bar{u_4}\gamma_5u_4$,
vector $\bar{u_4}\gamma_\mu u_4$ and axial vector $\bar{u_4}\gamma_5\gamma_\mu u_4$ are presented in figures (\ref{fig2}-\ref{fig5}) at three different fixed values from the considered working region for the continuum threshold.
From these figures, we see a good stability of the masses with respect to the Borel mass parameter $M^2$. From these figures, it is also clear that the results do not depend  on the continuum threshold in
its working region. The dependence of the decay constants of the  scalar $\bar{u_4}u_4$, pseudoscalar $\bar{u_4}\gamma_5u_4$,
vector $\bar{u_4}\gamma_\mu u_4$ and axial vector $\bar{u_4}\gamma_5\gamma_\mu u_4$ are presented in figures (\ref{fig6}-\ref{fig9}) also at three different fixed values of the continuum threshold.
These figures
also depict approximately insensitivity of the results under variation of the Borel mass parameter in its working region. The results of decay constants also show very weak dependency on the continuum threshold in its working region.
From a similar way, we analyze the mass and decay constants of the cases when one of the quarks belong  to the heavy fourth generation and the other is ordinary bottom or charm quark. The numerical results deduced
 from the figures are collected in Tables I-VI for three different values of the $m_{u_4}$, namely $m_{u_4}=450~ GeV$, $m_{u_4}=500~ GeV$ and $m_{u_4}=550~ GeV$. 
The errors presented in these tables are only due to the uncertainties coming from  determination of the working regions for the auxiliary parameters.
Here, we should stress that the obtained results in Tables I-VI are within QCD and do not include  contributions coming from the Higgs couplings to the ultra heavy quarks.
 Such contributions to the binding energy 
 have been calculated in  \cite{Ishiwata}, where it is shown that these contributions are more than several $GeV$ in the case when 
both quarks belong to the fourth family. The Higgs contribution calculated in \cite{Ishiwata} is proportional to the product of two quark masses. 
When we replace one of the ultra heavy quarks by $b$ or $c$ quark,
the binding energy obtained in \cite{Ishiwata} reduces to a value which is less than the QCD sum rules predictions in the present work. However, when both quarks belong to the fourth family, the binding energy 
obtained in the present work
is very small comparing to the Higgs corrections in \cite{Ishiwata}.

At the end of this part, we would like to mention that  the obtained QCD sum rules in the present work reproduce the masses and decay constants of the ordinary $\bar b b(\bar c c)$ states when we set $u_4\rightarrow b(c)$. The obtained numerical
 values in this limit are in a good consistency with the existing experimental data \cite{pdg} and QCD sum rules predictions \cite{Kazem3,Kazem4}.

To sum up, against the  top quark, the heavy  fourth generation of quarks that  have sufficiently small mixing with the three known  SM  families form hadrons. Considering the arguments mentioned in the text,
 the production
of such bound states  will be possible at LHC. Hoping for this  possibility, we calculated the masses and decay constants
of the bound state
 objects containing two  quarks  either both from the SM$_4$ or one from heavy fourth generation and the other from observed SM bottom or charm quarks in the framework of the QCD
 sum rules. The obtained numerical results  approach to the known masses and decay constants of the
$\bar b b$ and $\bar c c$ heavy quarkonia, when the fourth family quark is replaced by the bottom or charm quark.

%%%%%%%%%%%%%%%%%%%%%%%%%%%%%%%%%%%%%%%%%%%%%%%%%%%%%%%%%%%%%%%%%%%%%%%%%%%%%%

\begin{table}
\renewcommand{\arraystretch}{1.5}
\addtolength{\arraycolsep}{3pt}4
$$
\begin{array}{|c|c|c|c|c|}
\hline \mbox{mass (GeV)} & u_4\bar c  &u_4\bar b  & u_4\bar{u}_4    \\
\hline  \mbox{Scalar}          &453.01\pm0. 25 &456.45\pm0. 25 & 901.68 \pm0.50 \\
\hline   \mbox{Pseudoscalar}   &  452.62\pm0.15& 455.95 \pm0.15&  901.12 \pm0.30 \\
\hline  \mbox{axial vector}    &  453.00\pm0.25&  456.44\pm0.25&  901.70\pm0.50 \\
\hline  \mbox{vector}         & 452.62\pm0.15 & 455.94\pm0.15 &  901.13\pm0.30  \\
\hline

\end{array}
$$
\caption{The values of masses of different bound states obtained using $m_{u_4}=450~GeV$.}
\renewcommand{\arraystretch}{1}
\addtolength{\arraycolsep}{-3pt}
\end{table}

\begin{table}
\renewcommand{\arraystretch}{1.5}
\addtolength{\arraycolsep}{3pt}
$$
\begin{array}{|c|c|c|c|c|}
\hline \mbox{mass (GeV)} & u_4\bar c  &u_4\bar b  & u_4\bar{u}_4    \\
\hline  \mbox{Scalar}       &502.91\pm0.28    &506.36\pm0.28  & 1001.61\pm0.55  \\
\hline   \mbox{Pseudoscalar}&  502.52\pm0. 17  & 505.86\pm0.17 &  1001.04\pm0. 33 \\
\hline  \mbox{Axial Vector}  &  502.91\pm0.28  &  506.35\pm0.28&  1001.60\pm0.55 \\
\hline  \mbox{Vector}        & 502.57\pm0.  17  & 505.85\pm0.17 &  1001.04\pm0. 33 \\
\hline

\end{array}
$$
\caption{The values of masses of different bound states obtained using $m_{u_4}=500~GeV$.}
\renewcommand{\arraystretch}{1}
\addtolength{\arraycolsep}{-3pt}
\end{table}

\begin{table}
\renewcommand{\arraystretch}{1.5}
\addtolength{\arraycolsep}{3pt}
$$
\begin{array}{|c|c|c|c|c|}
\hline \mbox{mass (GeV)} & u_4\bar c  &u_4\bar b  & u_4\bar{u}_4    \\
\hline  \mbox{Scalar}          &552.82 \pm0.31 &556.27\pm0. 31 & 1101.67\pm0.60  \\
\hline   \mbox{Pseudoscalar}   & 552.43 \pm0.18& 555.78\pm0.18 &  1101.11\pm0.36  \\
\hline  \mbox{Axial Vector}    & 552.81 \pm0.31& 556.25\pm0.31&  1101.68\pm0.60 \\
\hline  \mbox{Vector}          & 552.42\pm0.18 & 555.77 \pm0.18&  1101.12 \pm0. 36\\
\hline

\end{array}
$$
\caption{The values of masses of different bound states obtained using $m_{u_4}=550~GeV$.}
\renewcommand{\arraystretch}{1}
\addtolength{\arraycolsep}{-3pt}
\end{table}

\begin{table}
\renewcommand{\arraystretch}{1.5}
\addtolength{\arraycolsep}{3pt}
$$
\begin{array}{|c|c|c|c|c|}
\hline \mbox{Leptonic decay constant f (GeV)} & u_4\bar c  &u_4\bar b  & u_4\bar{u}_4    \\
\hline  \mbox{Scalar}          &0.12 \pm0.01&0.15\pm0.02 & 0.28\pm0.03 \\
\hline   \mbox{Pseudoscalar}   & 0.17\pm0.01& 0.34\pm0.02& 4.01 \pm0.20\\
\hline  \mbox{Axial Vector}    & 0.12\pm0.01 & 0.15\pm0.02& 0.28\pm0.03 \\
\hline  \mbox{Vector}          &0.17\pm0.01&0.34\pm0.02 &  4.01\pm0.20 \\
\hline

\end{array}
$$
\caption{The values of decay constants of different bound states obtained using $m_{u_4}=450~GeV$.}
\renewcommand{\arraystretch}{1}
\addtolength{\arraycolsep}{-3pt}
\end{table}

\begin{table}
\renewcommand{\arraystretch}{1.5}
\addtolength{\arraycolsep}{3pt}
$$
\begin{array}{|c|c|c|c|c|}
\hline \mbox{Leptonic decay constant f (GeV)} & u_4\bar c  &u_4\bar b  & u_4\bar{u}_4    \\
\hline  \mbox{Scalar}          & 0.11\pm0.01 &0.13\pm0.01& 0.26\pm0.03 \\
\hline   \mbox{Pseudoscalar}   & 0.15\pm0.01& 0.30 \pm0.02& 3.91\pm0.19  \\
\hline  \mbox{Axial Vector}    & 0.11\pm0.01& 0.13\pm0.01&  0.26\pm0.03 \\
\hline  \mbox{Vector}          & 0.15\pm0.01&0.29\pm0. 02&  3.91\pm0.19 \\
\hline

\end{array}
$$
\caption{The values of decay constants of different bound states obtained using $m_{u_4}=500~GeV$.}
\renewcommand{\arraystretch}{1}
\addtolength{\arraycolsep}{-3pt}
\end{table}\

\begin{table}
\renewcommand{\arraystretch}{1.5}
\addtolength{\arraycolsep}{3pt}
$$
\begin{array}{|c|c|c|c|c|}
\hline \mbox{Leptonic decay constant f (GeV)} & u_4\bar c  &u_4\bar b  & u_4\bar{u}_4    \\
\hline  \mbox{Scalar}          &0.10\pm0.01&0.12\pm0.01 & 0.26\pm0.03 \\
\hline   \mbox{Pseudoscalar}   & 0.14\pm0.01& 0.27\pm0.01& 4.19\pm0.20  \\
\hline  \mbox{Axial Vector}    &0.10\pm0.01& 0.12\pm0.01&  0.26\pm0. 03\\
\hline  \mbox{Vector}          & 0.14 \pm0.01&0.27\pm0.01 &  4.18\pm0.20  \\
\hline

\end{array}
$$
\caption{The values of decay constants of different bound states obtained using $m_{u_4}=550~GeV$.}
\renewcommand{\arraystretch}{1}
\addtolength{\arraycolsep}{-3pt}
\end{table}

\end{document}